\documentclass[12pt]{article}
\usepackage{graphicx}
\usepackage[bookmarks=true,pdfborder={0 0 0}]{hyperref}

\newcommand\pubnumber{CMS CR-2014/149}
\newcommand\pubdate{\today}

\def\napoli{Institute of High Energy Physics,\\
Chinese Academy of Science}

\def\support{\footnote{Work supported in part by the Ministry of Science Technology of China, 973 project and National Natural Science Foundation of China.}}
\def\Title#1{\begin{center} {\Large #1 } \end{center}}
\def\Author#1{\begin{center}{ \sc #1} \end{center}}
\def\Address#1{\begin{center}{ \it #1} \end{center}}

\newcommand\pubblock{\rightline{\begin{tabular}{l} \pubnumber\\
         \pubdate  \end{tabular}}}
\newenvironment{Abstract}{\begin{quotation}  }{\end{quotation}}
\newenvironment{Presented}{\begin{quotation} \begin{center}
             PRESENTED AT\end{center}\bigskip
      \begin{center}\begin{large}}{\end{large}\end{center} \end{quotation}}
\def\Acknowledgements{\bigskip  \bigskip \begin{center} \begin{large}
             \bf ACKNOWLEDGEMENTS \end{large}\end{center}}

\def\beq{\begin{equation}}
\def\eeq#1{\label{#1}\end{equation}}
\def\eeqn{\end{equation}}

\def\beqa{\begin{eqnarray}}
\def\eeqa#1{\label{#1}\end{eqnarray}}
\def\eeqan{\end{eqnarray}}

\let\bar=\overbar

\def\Dslash{\not{\hbox{\kern-4pt $D$}}}
\def\dslash{\not{\hbox{\kern-2pt $\del$}}}

\def\msb{{\bar{\ssstyle M \kern -1pt S}}}

\begin{document}
\begin{titlepage}
\pubblock

\vfill
\Title{Exotic hadrons at hadron colliders}
\vfill
\Author{Ye Chen\support \it{on behalf of the CMS collaboration}. \\ \it{This talk also contains material of LHCb, D0 and CDF.} }
\Address{\napoli}
\vfill
\begin{Abstract}
In this proceeding, an overview of the recent progress of the exotic hadrons studies at hadron colliders is presented, including
the experimental measurement results from CMS, LHCb, CDF and D0. The talk covers the physics properties study of X(3872); the
search for Y(4140) state etc; the recent result of Z(4430); and also the extended study to bottomonium sector.

\end{Abstract}

\begin{Presented}
2014 Flavor Physics and CP Violation (FPCP-2014), Marseille, France, May 26-30 2014, 9 pages, 9 figures.
\end{Presented}
\vfill
\end{titlepage}
\def\thefootnote{\fnsymbol{footnote}}
\setcounter{footnote}{0}

\section{Introduction}
the exotic hadrons states, which have candidates known as XYZ family, have rich research results in recent decades. The discovery of X(3872) particle in 2003~\cite{belle2003} starts the stories of the XYZ research. About twenty candidates of the exotic hadrons states have been appeared both at $e^{+}e^{-}$ colliders and hadron colliders. As most of those candidate states do not fit into conventional quark model predictions, lots of research interests among hadron physics have been attracted. The possible explanations for those states are the multi-quark states, molecular states, hybrid meson and glueball etc.

In this talk, an overview of the recent progress of the exotic hadrons studies at hadron colliders is presented, including the experimental measurement results from CMS, LHCb, CDF and D0.
\section{Recent Studies of X(3872) states}

\begin{figure}[htb]
\centering
\includegraphics[height=2.in]{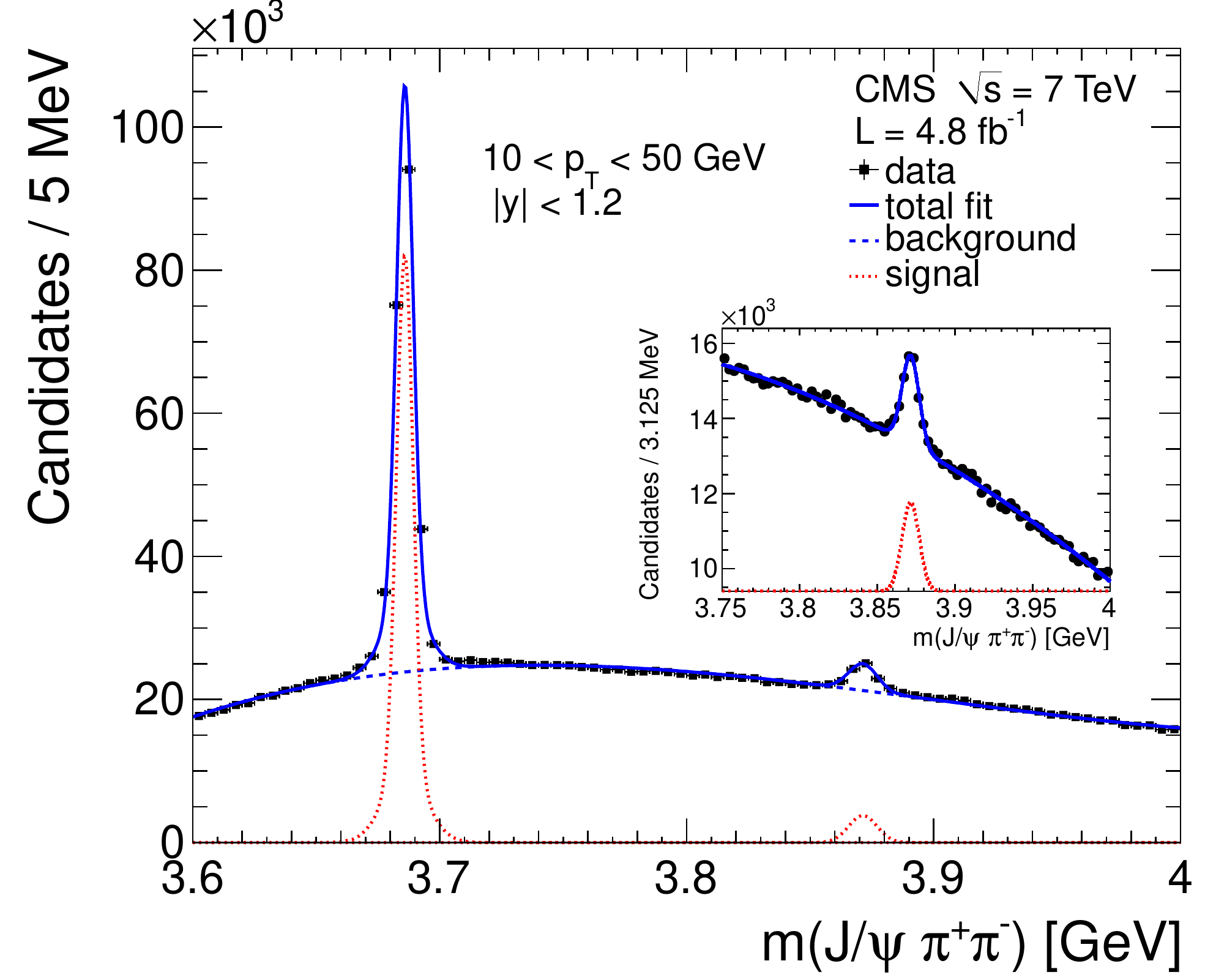}
\includegraphics[height=2.in]{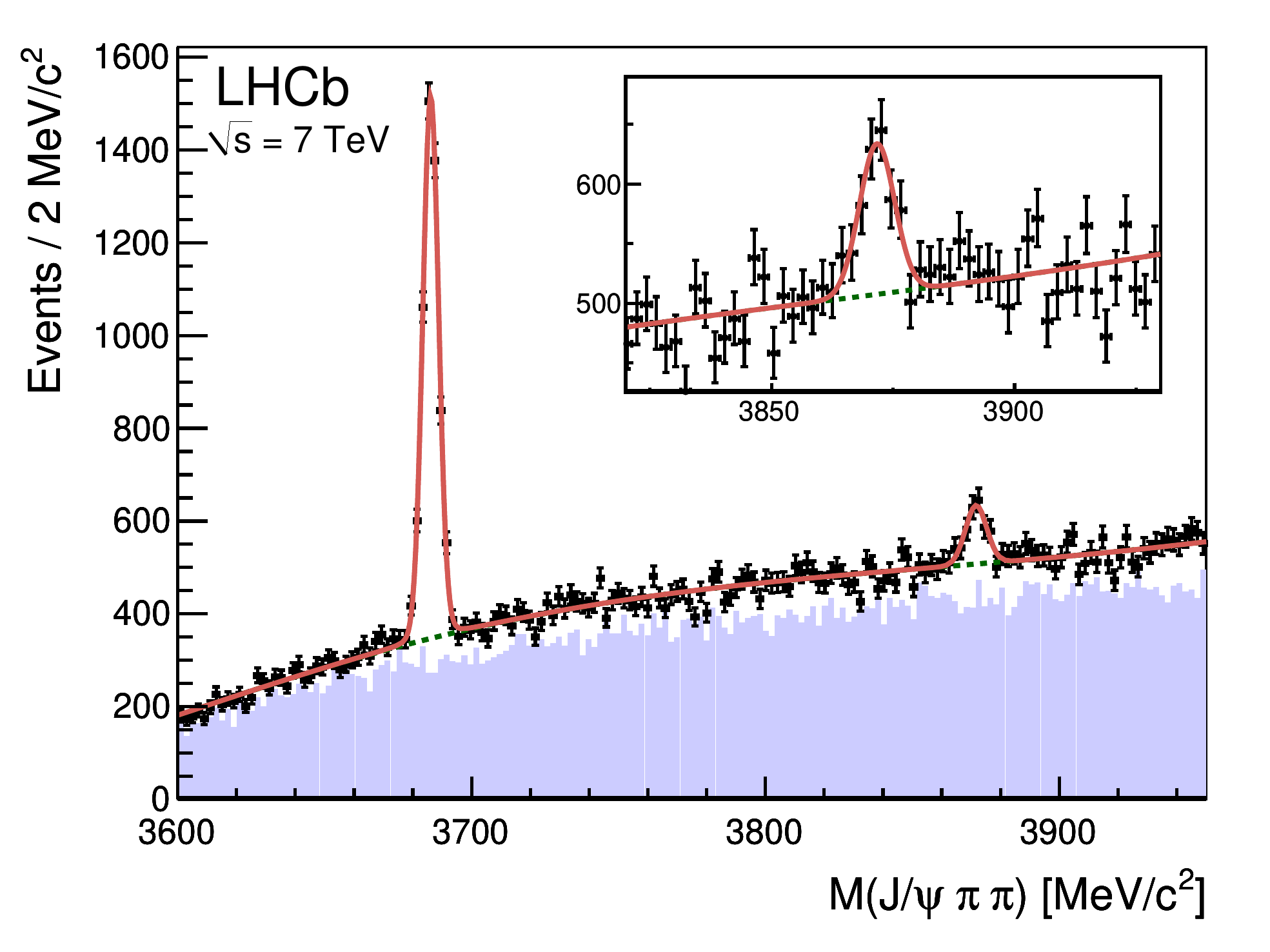}
\caption{Invariant mass spectrum of $J/\psi \pi \pi$ at Large Hadron Collider~\cite{cms2013,lhcb2012}.}
\label{fig:X3872}
\end{figure}

The X(3782) is the first exotic meson candidate discovered by BELLE in 2003~\cite{belle2003}. It is later confirmed by CDF~\cite{cdf2004}, D0~\cite{d02004} and Babar~\cite{babar2005}. The X(3872) also has been observed at Large Hadron Collider in both CMS experiment~\cite{cms2013} and LHCb experiment~\cite{lhcb2012}. Fig.~\ref{fig:X3872} shows the X(3872) signals at LHC experiment. The left plot is the X(3872) signal at the center rapidity region in CMS experiment~\cite{cms2013}. The right plot is the X(3872) signal at the forward rapidity region in LHCb experiment~\cite{lhcb2012}. The physics properties of X(3872) have also been investigated further with the date accumulated at LHC~\cite{cms2013,lhcb2012,lhcb2014,lhcb2013}.

\begin{figure}[htb]
\centering
\includegraphics[height=2in]{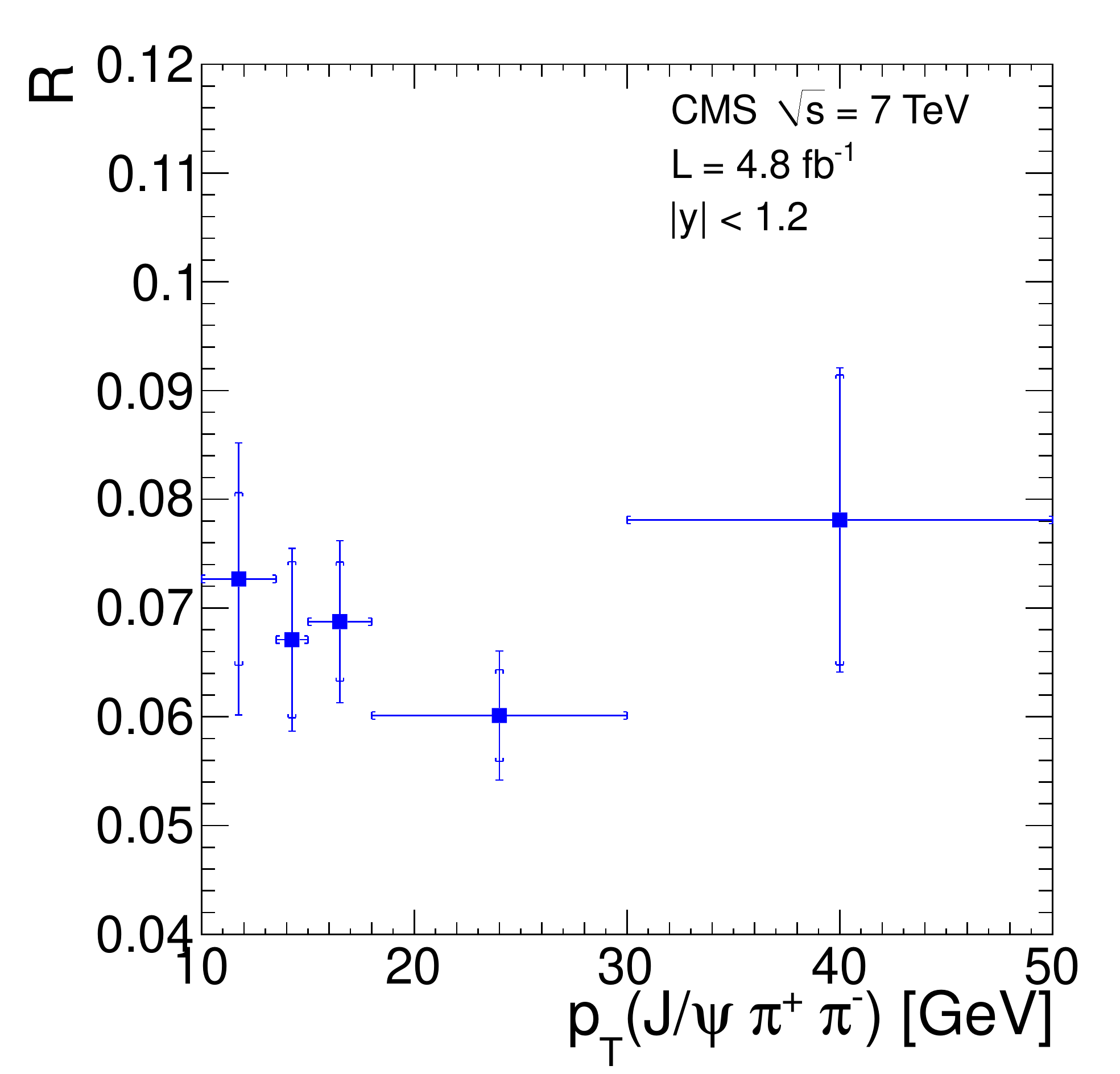}
\includegraphics[height=2in]{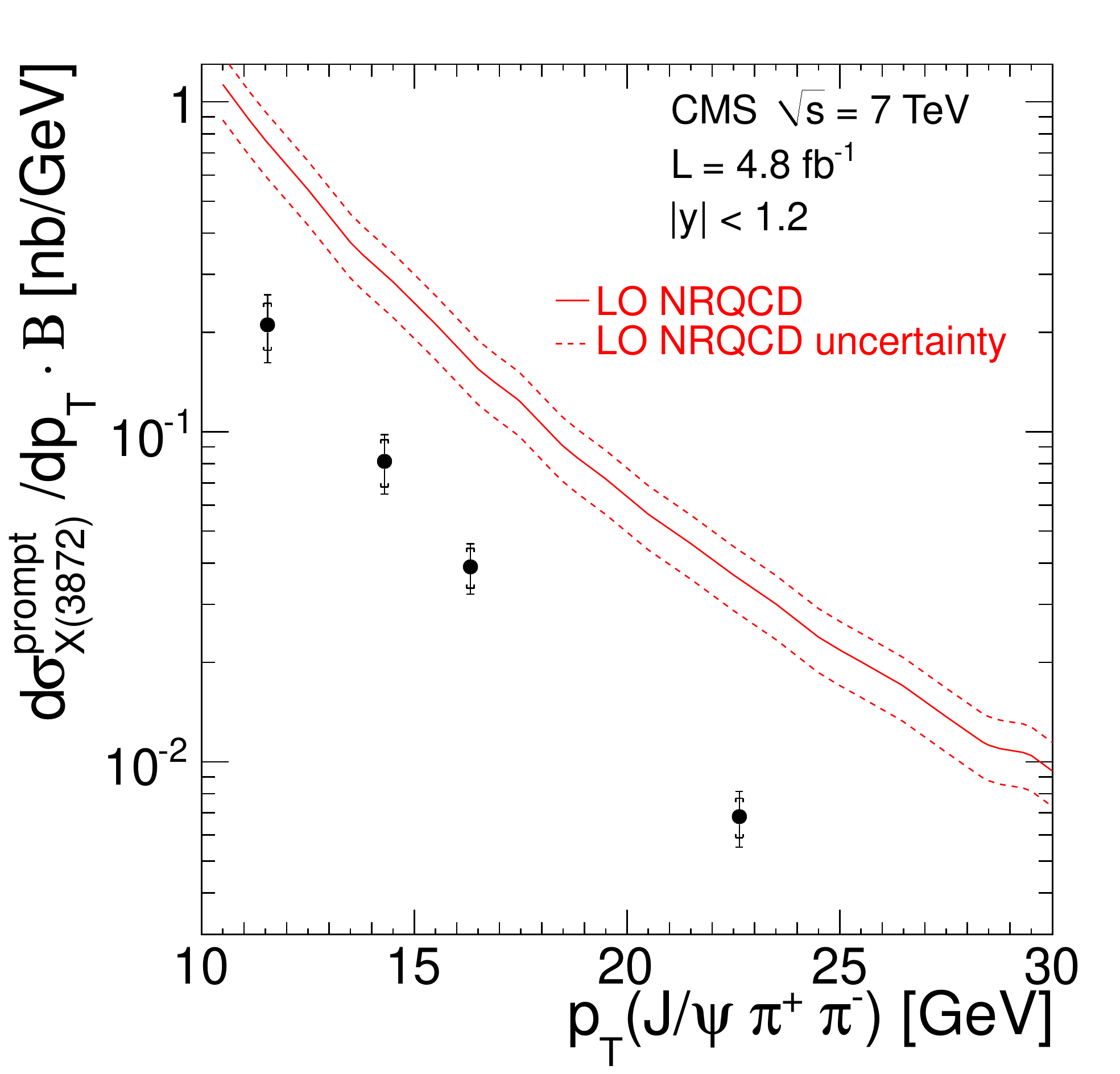}
\caption{Ratio of X(3872) and $\psi(2S)$ cross section (left), and prompt X(3872) $p_T$ differential cross section (right) as measured by the CMS experiment at LHC in pp collisions at $\sqrt{(s)}= 7 TeV$~\cite{cms2013}.}
\label{fig:cmsX38722}
\end{figure}

\begin{figure}[htb]
\centering
\includegraphics[height=1.8in]{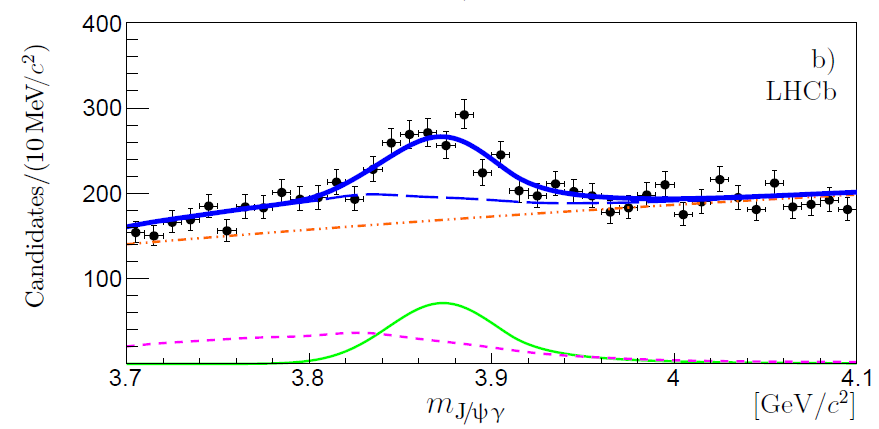} \\
\includegraphics[height=1.8in]{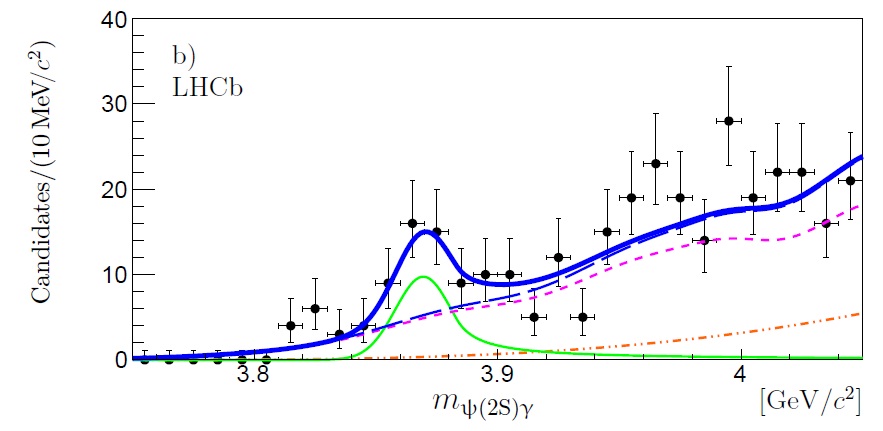}
\caption{Comparison of the $J/\psi \gamma$ invariant mass spectrum and the $\psi \gamma$ invariant mass spectrum. The total
fit (thick solid blue) together with the signal (thin solid green) and background components (dash-dotted orange for the
combinatorial, dashed magenta for the peaking component and long dashed blue for their sum) are shown~\cite{lhcb2014}.
}
\label{fig:lhcb1}
\end{figure}

In the 2013, CMS experiment did the measurement of X(3872)'s cross section~\cite{cms2013} presented in the left plot of Fig.~\ref{fig:cmsX38722}. The fraction of non-prompt X(3872) from B decays is measured as well. In the right plot of Fig.~\ref{fig:cmsX38722}, a comparison of the theoretical prediction and the experiment measured cross section of prompt X(3872) is given. The predicted NRQCD is above the measured cross section~\cite{cms2013}. Very recently, LHCb experiment has gotten an evidence of the decay of $X(3872) \rightarrow \gamma \psi(2S)$ with a significance of 4.4 $\sigma$~\cite{lhcb2014}. Fig.~\ref{fig:lhcb1} shows the $\gamma J/\psi$ and the $\gamma \psi(2S)$ invariant mass spectrum. The relative branching fraction measurement agrees with the expectation for a pure charmonium interpretation and the molecular-charmonium mixture interpretations while a pure $D \bar{D^{*}}$ molecular interpretation is not supported~\cite{lhcb2014}.

\begin{figure}[htb]
\centering
\includegraphics[height=1.5in]{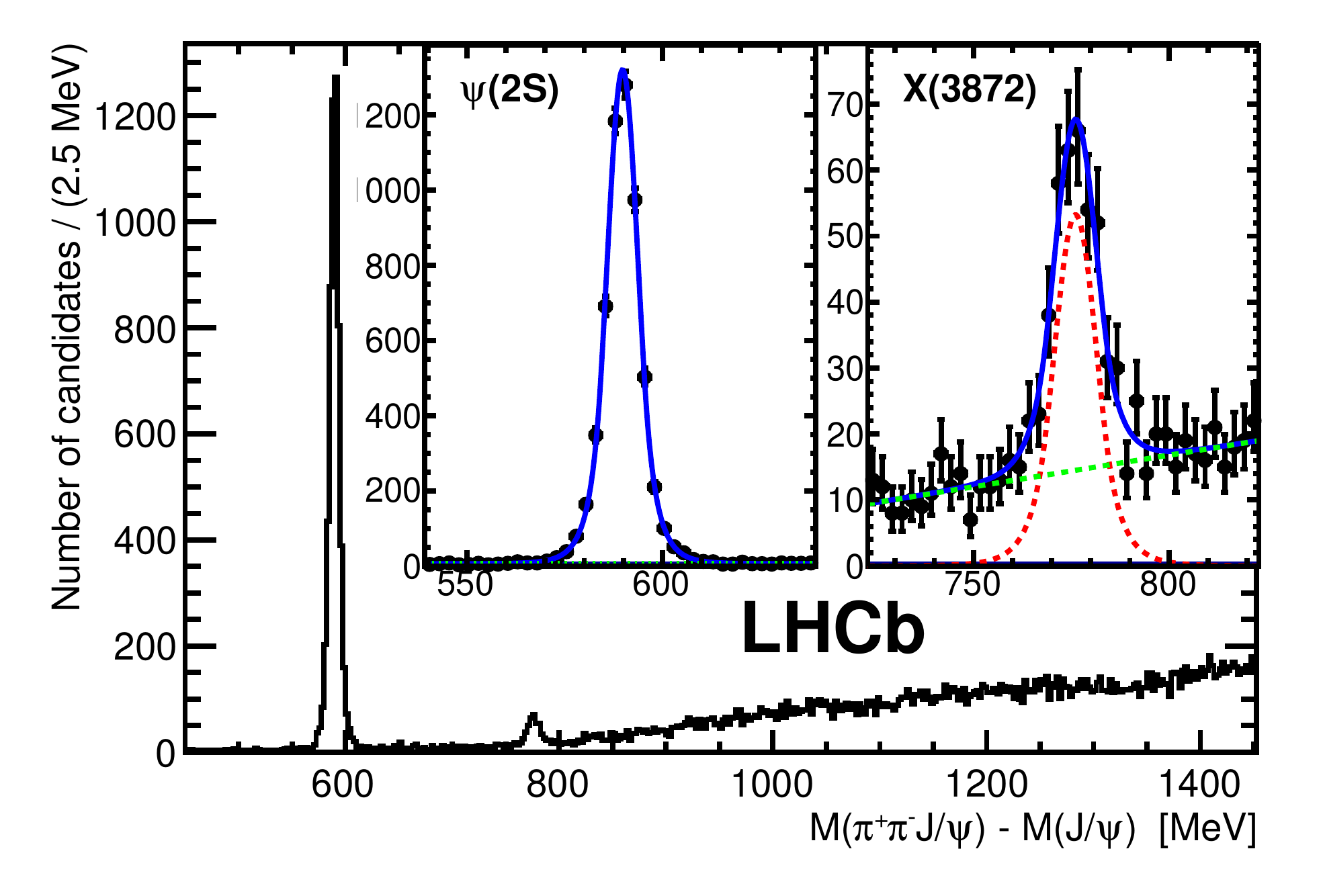}
\includegraphics[height=1.5in]{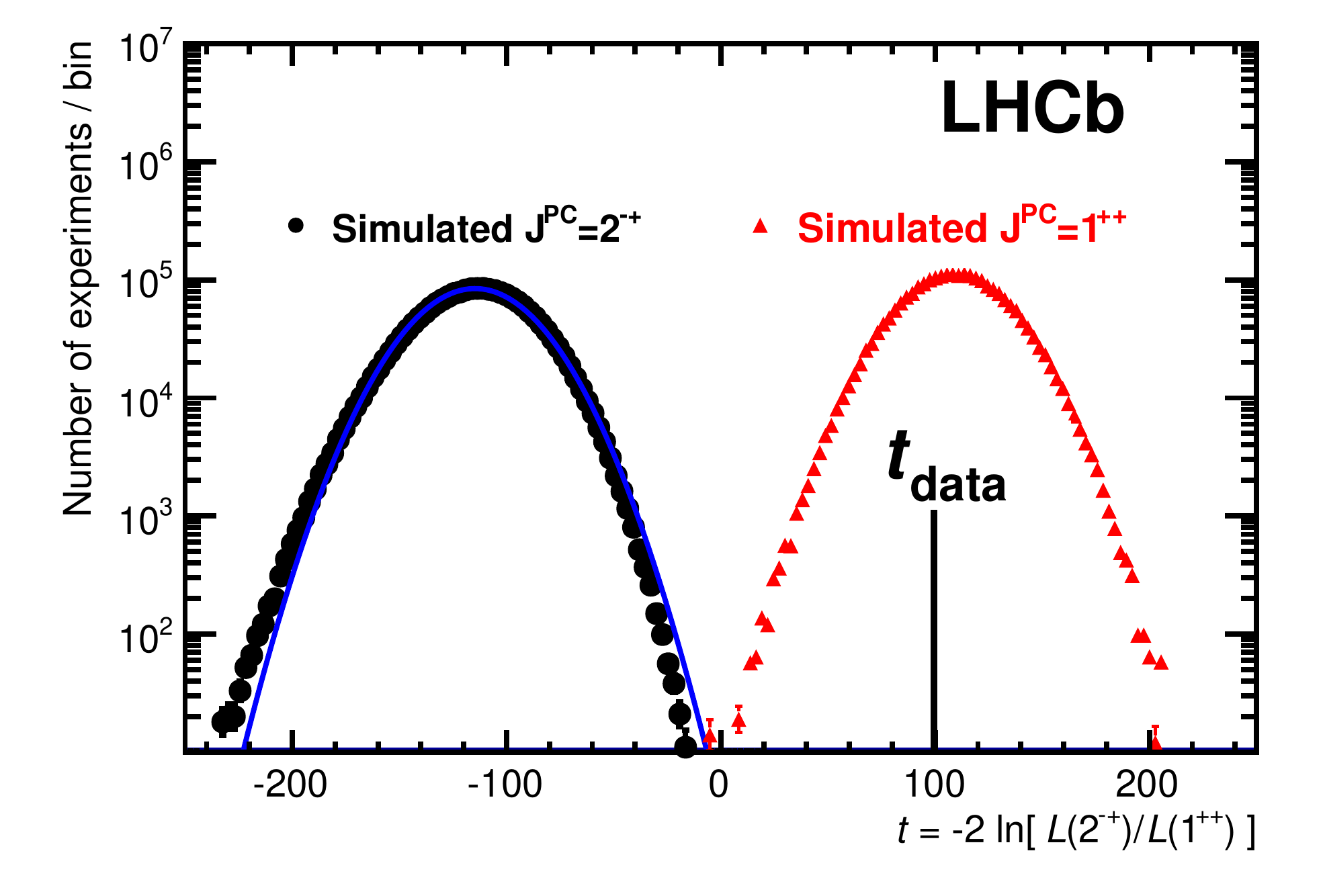}
\caption{X(3872) signal at LHCb and the $J^{pc}$ determination~\cite{lhcb2013}.}
\label{fig:lhcb2}
\end{figure}

The $J^{pc}$ of the X(3872) has been recently measured by the LHCb experiment as well and is determined to be $1^{++}$~\cite{lhcb2013}. The only allowed alternative $J^{pc}$ of X(3872) is $2^{++}$ according to previous measurement. From the right plot of Fig.~\ref{fig:lhcb2}, the $2^{++}$ hypothesis is rejected with a confidence level equivalent to more than 8 standard deviations. This result favors exotic explanation of the X(3872) states~\cite{lhcb2013}.

\section{Search for Y(4140) and Y(4270)}

\begin{figure}[htb]
\centering
\includegraphics[height=2.0in]{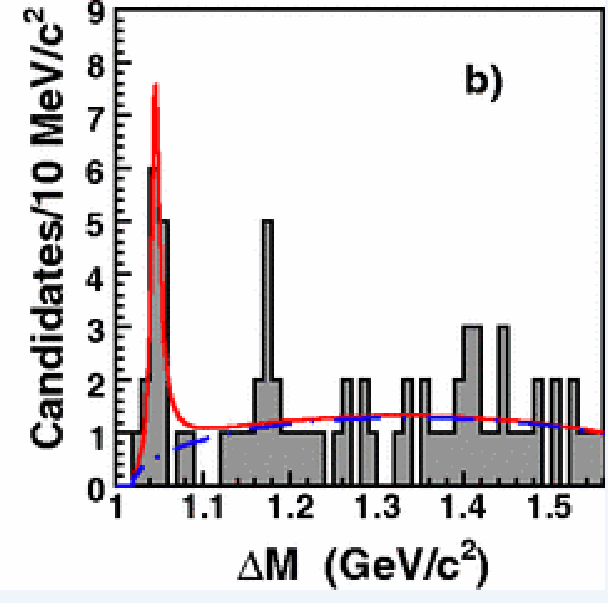}
\includegraphics[height=2.0in]{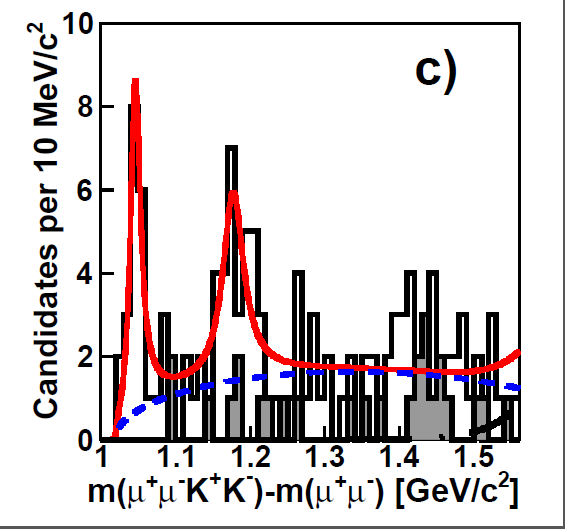}
\caption{ $\Delta M$ spectrum of $M(J/\psi \phi)-M(J/\psi)$ in CDF experiment~\cite{cdf2009,cdf2011}. The left plot corresponds to 2.7 $fb^{-1}$ and the right one corresponds to 6 $fb^{-1}$. }
\label{fig:cdf}
\end{figure}

Y(4140) has been observed firstly at CDF in 2009 using $2.7 fb^{-1}$ of data in the mass spectrum of $J/\psi \phi$, and a significance of $3.8 \sigma$~\cite{cdf2009}. Later with 6 $fb^{-1}$ of CDF data, the significance of Y(4140) is about $5 \sigma$ and an also an evidence for a second structure was reported~\cite{cdf2011}. The measurements results have been presented in Fig.~\ref{fig:cdf}.

\begin{figure}[htb]
\centering
\includegraphics[height=1.4in]{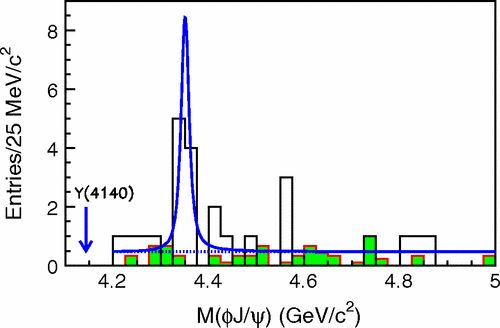} \\
\includegraphics[height=2.5in]{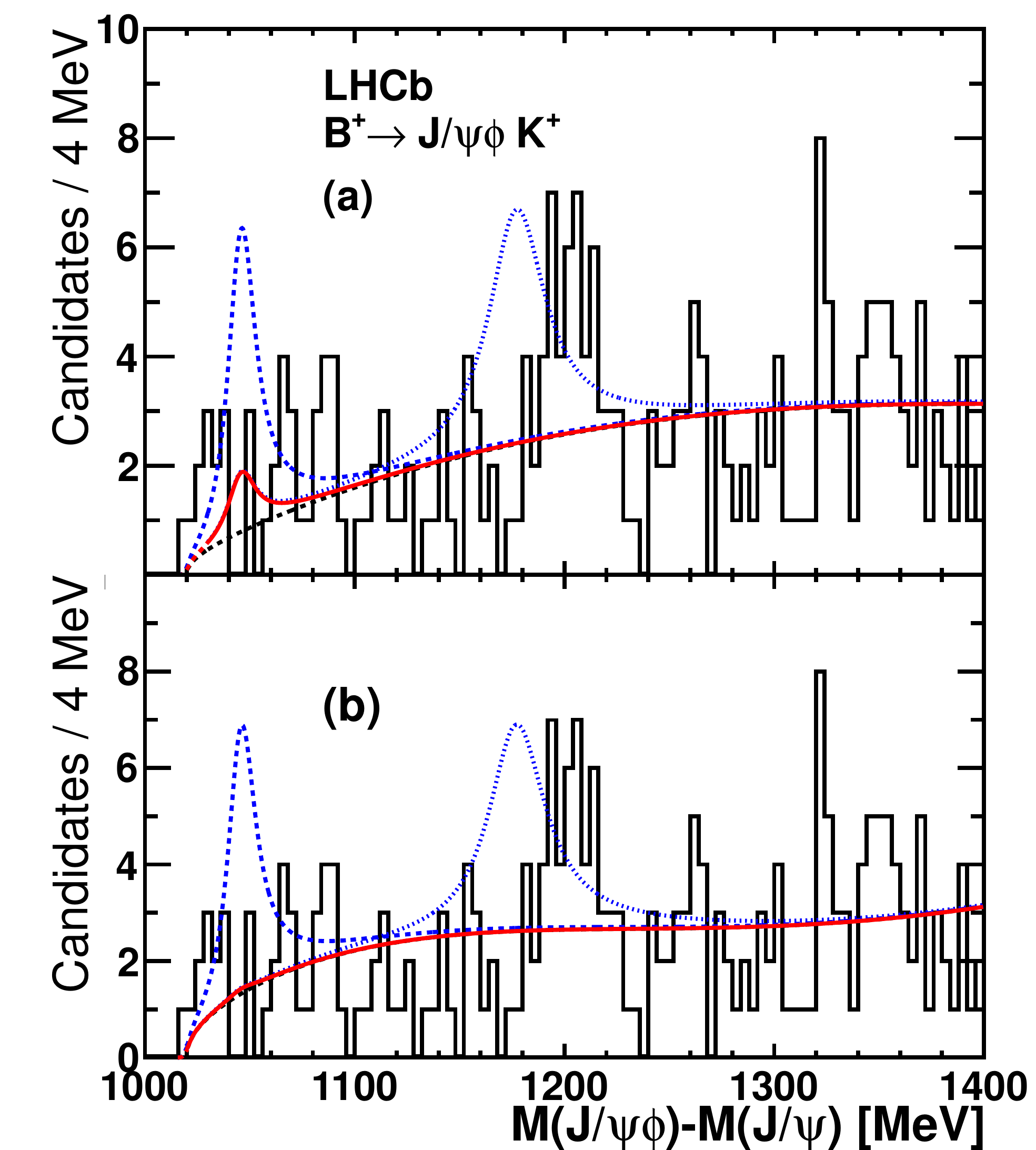}
\caption{The top plot is the invariant mass spectrum of $J/\psi \phi$ in Belle experiment~\cite{belle2010}. The middle and the bottom plots are the $\Delta M$ spectrum of $M(J/\psi \phi)-M(J/\psi)$ in LHCb experiment with two different background scenarios~\cite{lhcb2012B}.}
\label{fig:y4140_2}
\end{figure}

\begin{figure}[htb]
\centering
\includegraphics[height=2.8 in]{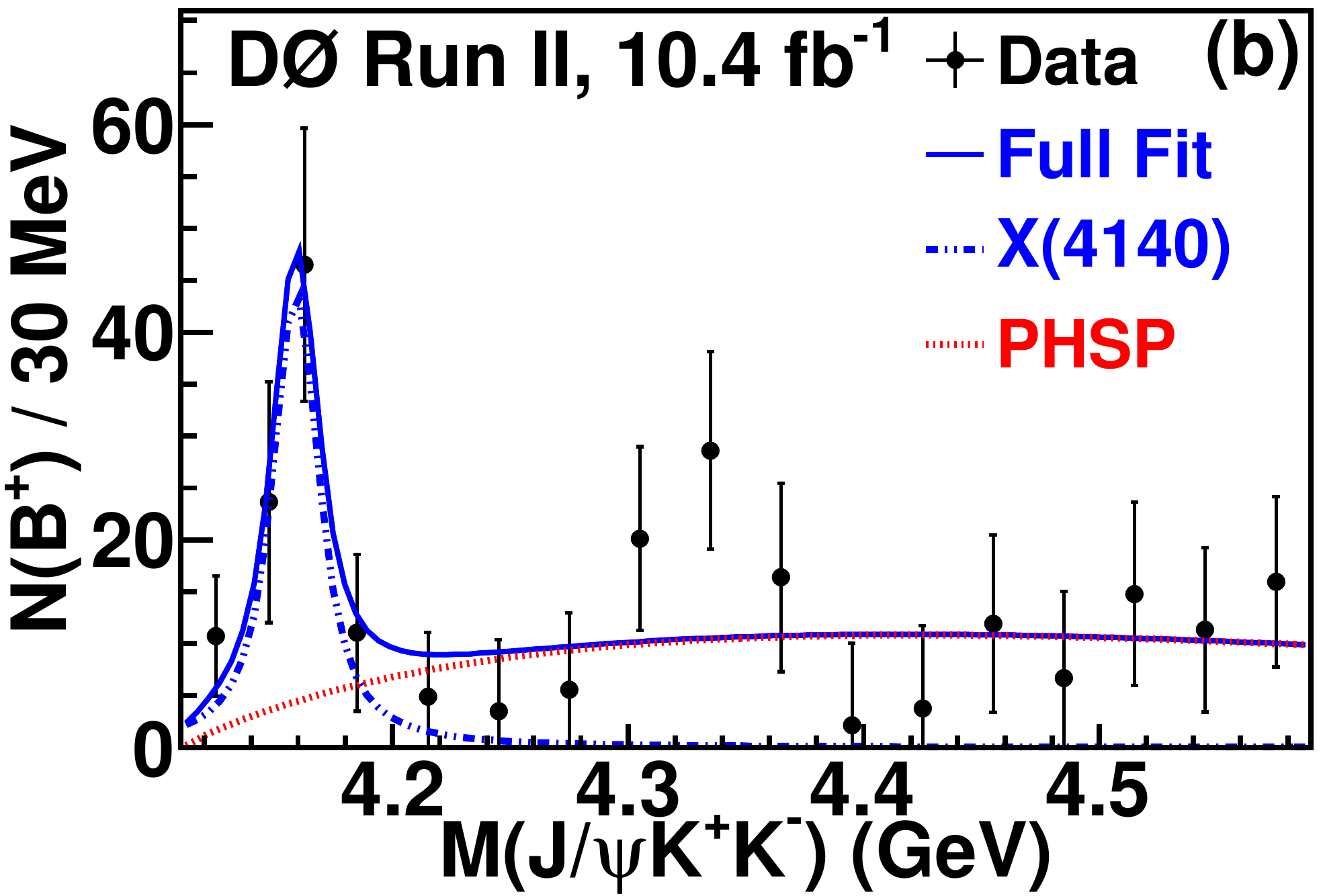} \\
\includegraphics[height=1.6in]{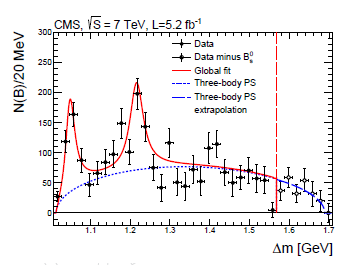}
\caption{The top and middle plots are the invariant mass spectrum in D0 experiment with two different fitting scenarios. The bottom plot is the $\Delta M$ spectrum of $M(J/\psi \phi)-M(J/\psi)$ in CMS experiment~\cite{d02014,cms2013B}.}
\label{fig:y4140_3}
\end{figure}

In the year of 2010, Belle also searched for Y(4140) in the $J/\psi \phi$ mass spectrum (see the top plot of Fig.~\ref{fig:y4140_2})~\cite{belle2010} but found no structure near the threshold region of $J/\psi \phi$, while had an evidence for the second structure with mass around 4350 $MeV/c^{2}$~\cite{belle2010}. In the year 2012, LHCb did the search again and found no significant excess in either case (see the middle and bottom plots of Fig.~\ref{fig:y4140_2})~\cite{lhcb2012B}. D0 performed the same search in the year 2014 and an evidence for the existence of near $J/\psi \phi$ threshold structure is reported with 3.1 standard deviations~\cite{d02014}. CMS also released the result of the $J/\psi \phi$ invariant mass spectrum structure search~\cite{cms2013B}, and a signal exceeding 5 $\sigma$ is observed. In both D0 and CMS results, the second structure around 4330 MeV is present~\cite{d02014,cms2013B}. Fig.~\ref{fig:y4140_3} shows the $J/\psi \phi$ invariant mass spectrum at D0 and CMS. From the results above, we see there are some inconsistencies among those measurements and the study of Y(4140), Y(4270) or other possible states might need to go further in future.

\section{Charged Exotic Hadron at Hadron Collider}
\begin{figure}[htb]
\centering
\includegraphics[height=1.5in]{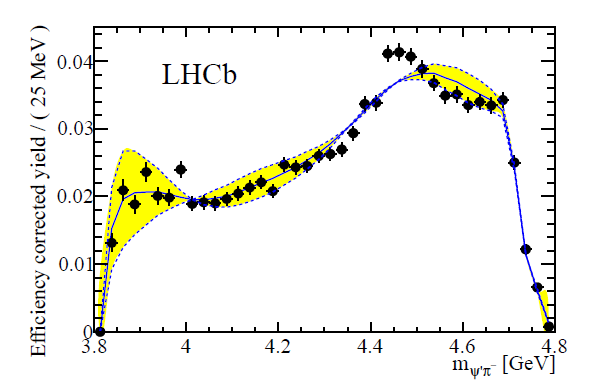}
\caption{Invariant mass spectrum of $\psi \pi$ at LHCb experiment~\cite{lhcb2014B}.}
\label{fig:lhcb4}
\end{figure}

The charged exotic hadrons are natural candidates for the four-quark state matter. There are a series of results of charged exotic hadrons at $e^{+} e^{-}$ colliders which will be included by Dr. Kai Zhu in the same session of the conference. For the hadron collider experiment, LHCb have reported the observation of $Z(4430)$ very recently~\cite{lhcb2014B}. Fig.~\ref{fig:lhcb4} shows the invariant mass spectrum of $\phi \pi$. The Z(4430) is observed with high significance and its spin-parity is determined to be $1^{+}$. The measurement favors four quark bound state explanation of Z(4430)~\cite{lhcb2014B}.

\section{Extended to Bottomonium Sector}
\begin{figure}[htb]
\centering

\includegraphics[height=1.5in]{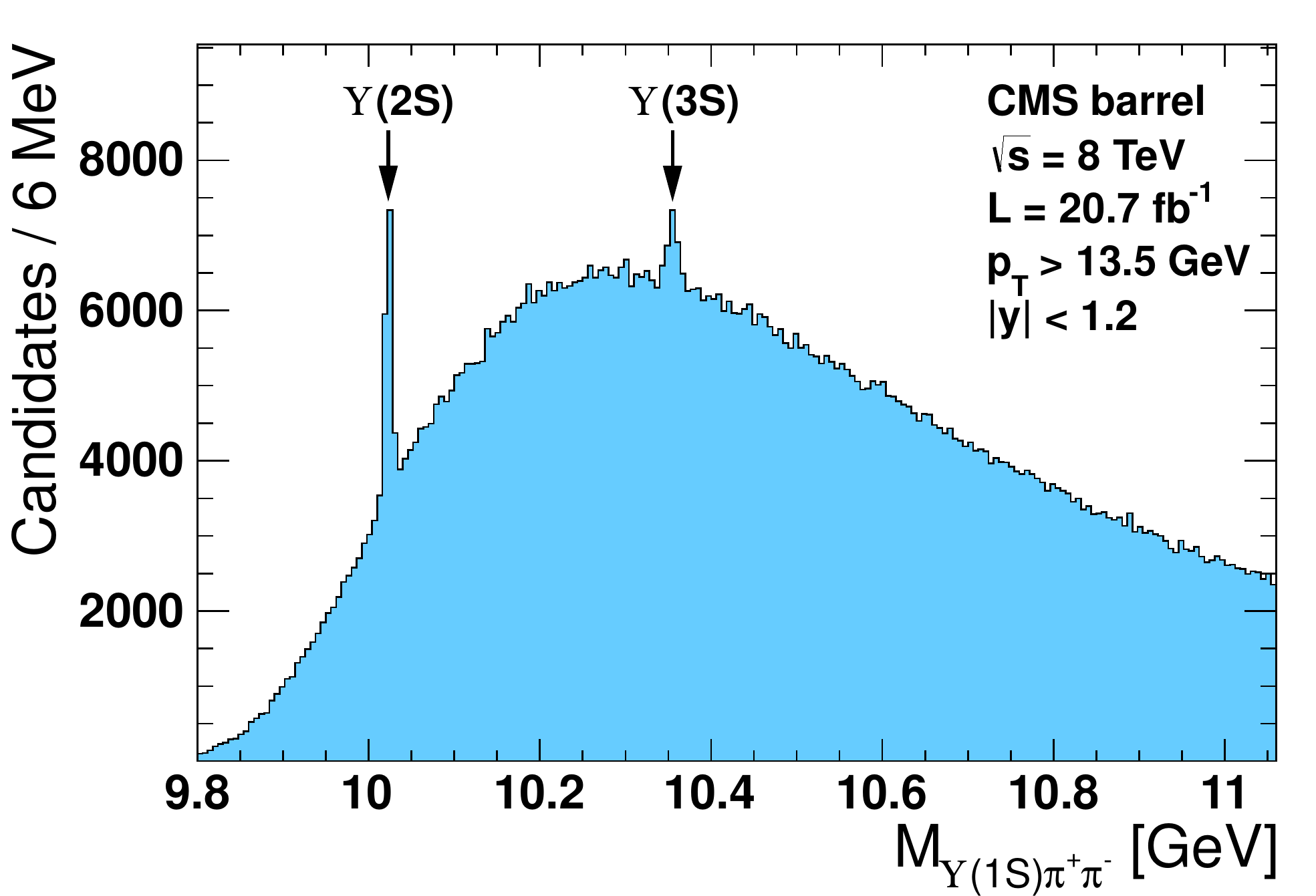}
\includegraphics[height=1.5in]{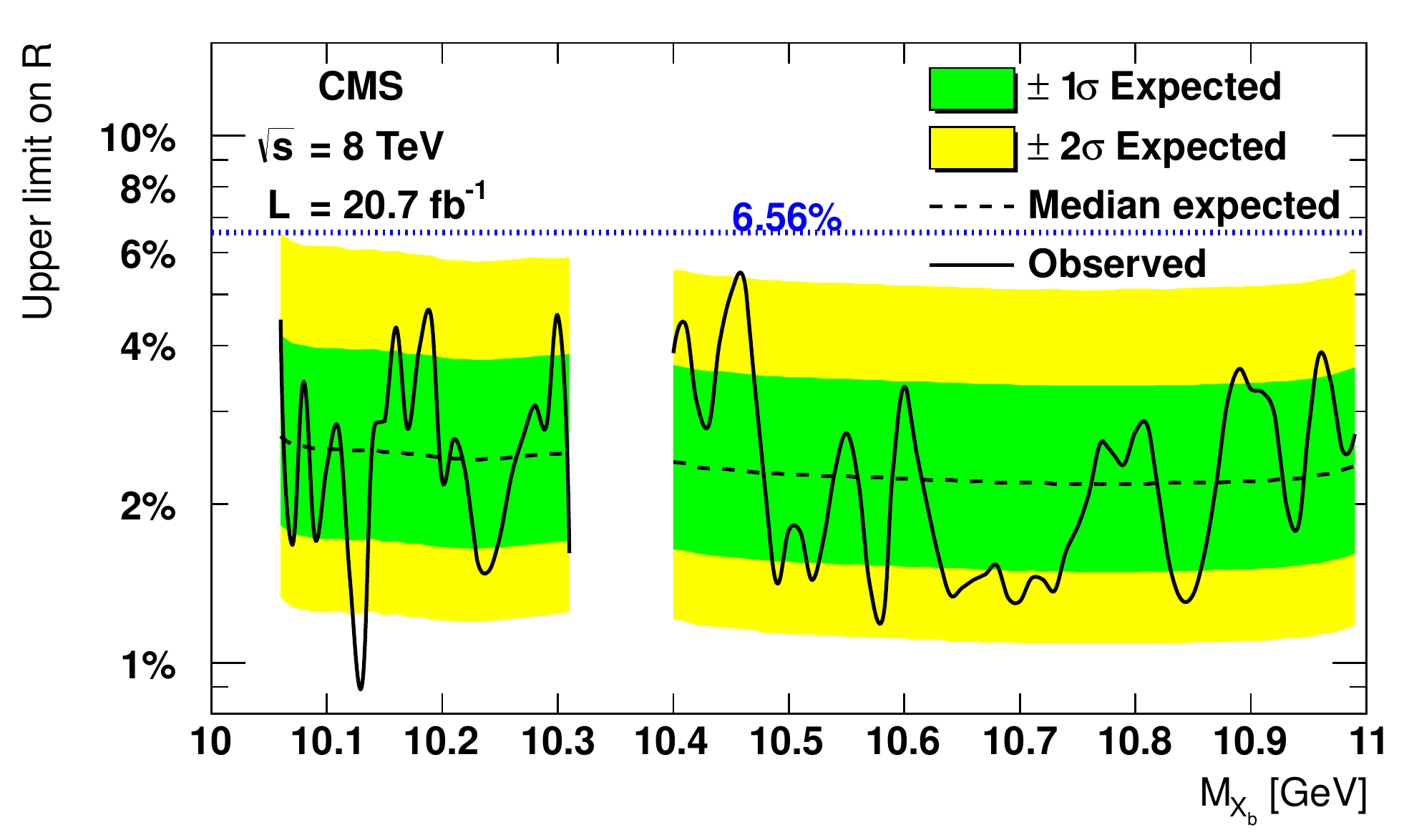}
\caption{Search for the bottomonium partner of X(3872) in the hadron collider~\cite{cms2013C}. The left plot is the invariant mass spectrum of $\Upsilon(1S) \pi^{+}\pi^{-}$ and the two peaks are the $\Upsilon(2S)$ and $\Upsilon(3S)$ signals respectively. The right plot is the upper limits at 95\% confidence level on the ratio of production cross section of the X$_b$ times its branching fraction to $\Upsilon(1S)\pi^{ +}\pi^{ −}$ , relative to the $\Upsilon(2S)$~\cite{cms2013C}.}
\label{fig:cms3}
\end{figure}

In the year of 2011, Belle has claimed the observations of two exotica states in $\Upsilon(5S)$ decays~\cite{belle2011}. Thus extending the XYZ exotic state family to the bottomonium sector. The CMS experiment also searched for the bottomonium partner of X(3872) at hadron colliders~\cite{cms2013C}. The search used the $\Upsilon(1S) \pi \pi$ decay mode and found no evidence for the X$_b$ state, as shown in the left plot of Fig.~\ref{fig:cms3}. The right plot of Fig.~\ref{fig:cms3} shows the ratio of the cross section X$_b$ to $\Upsilon(2S)$ upper limit is set in the range of (0.9-5.4)\% at 95\% confidence level for $X_b$ masses between 10 and 11 GeV. Those are the first upper limits on the production of a possible $X_b$ at a hadron collider.

\section{Summary and Outlook}
To summarize, rich results on exotic hadron states have been gotten in recent years at hadron colliders like CMS, LHCb and D0 experiments etc. This talk covers the most recent study on X(3872), Y(4140) , charged Z(4430) and extension to the bottomonium sector. For the study of X(3872), the spin-parity of X(3872) has been determined. The physics nature of X(3872) has been understood better with the data collected from the hadron collider. For the Y(4140), the evidence/observation of this exotic hadron state has been reported; the spin parity of Z(4430) has also been determined; and the search has been extended to bottomonium sector at hadron collider.

\Acknowledgements
Many thanks to FPCP 2014 LOC for holding this wonderful conference. The speaker is also grateful to the CMS BPH conveners Fabrizio Palla and Kai-Feng(Jack) Chen for helpful revisions. Many thanks to CMS conference committee for granting the talk opportunity. The speaker would also like to thank Kai Yi and BPH group in IHEP, CAS including Guoming Chen, Jianguo Bian, Zheng Wang and Xiangwei Meng for inspiration and helpful discussion during the cooperation.

The speaker is supported in part by National Natural Science Foundation of China under Contact No. 11061140514 and Ministry of Science and Technology of China, 973 project under Contact No. 2013B837801.


\begin{thebibliography}{99}
\bibitem{belle2003}
\href{http://dx.doi.org/10.1103/PhysRevLett.91.262001}{BELLE Collaboration, Phys. Rev. Lett. \bf{91},(2003).}

\bibitem{cdf2004}
\href{http://dx.doi.org/10.1103/PhysRevLett.93.072001}{CDF Collaboration, Phys. Rev. Lett. {\bf 93}, 072001 (2004).}

\bibitem{d02004}
\href{http://dx.doi.org/10.1103/PhysRevLett.93.162002}{D0 Collaboration, Phys. Rev. Lett. {\bf 93}, 162002 (2004).}


\bibitem{babar2005}
\href{http://dx.doi.org/10.1103/PhysRevD.71.071103}{BABAR Collaboration, Phys. Rev. \bf{D 71}, 071103 (2005).}


\bibitem{cms2013}
\href{http://dx.doi.org/10.1007/JHEP04(2013)154}{CMS Collaboration, JHEP {\bf 1304},154 (2013)}

\bibitem{lhcb2012}
\href{http://dx.doi.org/10.1140/epjc/s10052-012-1972-7}{LHCb Collaboration, Eur. Phys. J. {\bf C 72}, 1972 (2012)}

\bibitem{lhcb2014}
\href{http://arxiv.org/abs/arXiv:1404.0275}{LHCb Collaboration, arXiv:1404.0275 (2014)}

\bibitem{lhcb2013}
\href{http://dx.doi.org/10.1103/PhysRevLett.110.222001}{LHCb Collaboration, Phys. Rev. Lett. {\bf 110}, 211801 (2013)}

\bibitem{cdf2009}
\href{http://dx.doi.org/10.1103/PhysRevLett.102.242002}{CDF Collaboration, Phys. Rev. Lett.,{\bf 102}, 242002 (2009)}

\bibitem{cdf2011}
\href{http://arxiv.org/abs/arXiv:1101.6058}{CDF Collaboration, arXiv:1101.6058 (2011)}

\bibitem{belle2010}
\href{http://dx.doi.org/10.1103/PhysRevLett.104.112004}{Belle Collaboration, Phys. Rev. Lett, {\bf 104}, 112004 (2010)}

\bibitem{lhcb2012B}
\href{http://dx.doi.org/10.1103/PhysRevD.85.091103}{LHCb Collaboration, Phys. Rev. D {\bf 85}, 091103 (2012)}

\bibitem{d02014}
\href{http://dx.doi.org/10.1103/PhysRevD.89.012004}{D0 Collaboration, Phys. Rev. D {\bf 89}, 012004 (2014)}

\bibitem{cms2013B}
\href{http://arxiv.org/abs/arXiv:1309.6920}{CMS Collaboration, arXiv: 1309.6920 (2013)}

\bibitem{lhcb2014B}
\href{http://dx.doi.org/10.1103/PhysRevLett.112.222002}{LHCb Collaboration, arXiv:1404.1903(2014)}

\bibitem{belle2011}

\href{http://dx.doi.org/10.1103/PhysRevLett.108.122001}{Belle Collaboration,Phys. Rev. Lett. {\bf 108}, 122001 (2012)}

\bibitem{cms2013C}
\href{http://dx.doi.org/10.1016/j.physletb.2013.10.016}{CMS Collaboration, Phys. Lett. B {\bf 727}, 57-76 (2013)}



\end{thebibliography}
\end{document}